\documentclass{article}

\usepackage{arxiv}

\usepackage[utf8]{inputenc} 
\usepackage[T1]{fontenc}    

\usepackage{booktabs}       
\usepackage{nicefrac}       
\usepackage{microtype}      
\usepackage{lipsum}		
\usepackage{doi}

\usepackage{hyperref, url}
\usepackage{amsfonts, amsmath, amssymb, parskip, bbm}
\usepackage{xcolor, graphicx, tikz-cd, caption, subcaption}
\usepackage{algpseudocode, algorithm}
\usepackage{verbatim}
\usepackage{mathrsfs}

\usepackage{amsthm}  
\newtheorem{theorem}{Theorem}

\newtheorem{remark}{Remark}

\title{Constrained Dikin--Langevin diffusion for polyhedra}

\author{
  \hspace{1mm}James Chok \\
  School of Mathematics and \\
  Maxwell Institute for Mathematical Sciences, \\
  The University of Edinburgh,\\
  Edinburgh,\\
  EH9 3FD, \\
  United Kingdom\\
  \texttt{james.chok@ed.ac.uk} \\
    \And
  Domenic Petzinna \\
  Department of Mathematics,\\
  King's College London,\\
  London,\\
  WC2R 2LS,\\
  United Kingdom\\
  \texttt{domenicpetzinna12@gmail.com} \\
}

\renewcommand{\shorttitle}{Constrained Dikin--Langevin diffusion for polyhedrons}

\hypersetup{
pdftitle={Constrained Dikin--Langevin diffusion for polyhedra},
pdfsubject={math.NA},
pdfauthor={James Chok, Domenic Petzinna},
pdfkeywords={First keyword, Second keyword, More},
}

\begin{document}
\maketitle

\begin{abstract}
We propose a reflection-free Langevin framework for sampling and optimization on compact polyhedra. The method is based on the inverse Hessian of the logarithmic barrier, which defines a Dikin--Langevin diffusion whose drift and noise adapt to the local interior-point geometry. We show that trajectories started in the interior remain feasible for all finite times almost surely, so the constrained domain is preserved without reflections or projections. For computation, we discretize the diffusion using the Euler--Maruyama scheme and apply a Metropolis--Hastings correction, yielding a sampler that targets the exact constrained distribution. We also propose an annealed interacting variant for nonconvex optimization. Numerically, the Metropolis-adjusted method outperforms both the Dikin random walk and standard MALA on anisotropic box-constrained Gaussians, and the interacting optimizer escapes suboptimal basins more reliably than the non-interacting method.
\end{abstract}


\section{Introduction}
Constrained optimization over polyhedral domains arises in many areas of applied mathematics \cite{Adcock2025, BIGGS1978, FLETCHER1971, ONeill2020}, ranging from operations research \cite{Chen2024, Sharma2025, Sen2025} to engineering design \cite{Kovtunenko2006, Hintermuller2009, Baravdish2023}. We consider the problem of minimizing a function $f(x)$ over a compact polyhedron $U \subset \mathbb{R}^d$ defined by $K$ linear inequalities:
\begin{equation}\label{eq:optimization_problem}
    \min_{x\in U}f(x),\qquad\text{where}\qquad  U\ =\ \{x\ :\ a_i \cdot x\ \leq\ b_i;\ i=1,\ldots, K\},
\end{equation}
where $a_i\in\mathbb{R}^d$ and $b_i\in\mathbb{R}$. We assume that $f$ is $C^2(\Omega)$, where $\Omega$ is an open neighborhood of $U$ (so $\nabla f$ has locally Lipschitz-continuous gradients). As such, $f$ can be nonconvex over $U$, making optimization particularly challenging for deterministic methods like gradient descent.

Stochastic approaches offer a way to escape local optima by exploring the landscape of $f$. Rooted in simulated annealing \cite{Kirkpatrick1983, ern1985} and MCMC \cite{Robert2004} methods, a common strategy is to reformulate the optimization as a sampling problem. Specifically, one considers the distribution supported on $U$:
\begin{equation}
    \rho_{\beta}(x)\ =\ \frac{1}{Z_{\beta}}\mathbbm{1}_U(x)\exp(-\beta f(x)),\quad\text{where}\quad Z_{\beta}\ =\ \int_{U}\exp(-\beta f(x))\, dx,
\end{equation}
for some $\beta>0$ known as the inverse temperature, and $\rho_{\beta}$ is known as the Boltzmann distribution. In the low-temperature limit $\beta \to \infty$, $\rho_\beta(x)$ concentrates around the global minimizer(s) of $f$. Sampling from $\rho_\beta$, for large $\beta$, thus provides an approach to approximate global optima: one obtains diversified candidates near the best minima (which can then be refined by local optimization routines) with convergence, in probability, to the global minimum as $\beta \to \infty$ \cite{Holley1988, Cox2008}.

Directly sampling from $\rho_\beta$ is typically intractable when $d$ is large or $f$ is complex. Instead, one constructs a Markov chain whose invariant distribution is $\rho_\beta$. Two broad classes of such methods exist: Metropolis--Hastings algorithms \cite{Hastings1970, Metropolis1953} (discrete-time) and Langevin diffusions \cite{Langevin1908, Lemons1997} (continuous-time). In the unconstrained case $U=\mathbb{R}^d$, both approaches can achieve rapid convergence under mild conditions (e.g., log-concavity or certain smoothness conditions), with exponential convergence in some settings \cite{Pavliotis2014, Livingstone2021, roberts_1996_mala, roberts_1996_aperiodic, BouRabee2009}.

{This paper first formulates the overdamped Dikin--Langevin diffusion, driven by the inverse log-barrier Hessian, and proves that its trajectories remain inside $U$ for all finite times almost surely while targeting the constrained Boltzmann distribution. We next derive a Metropolis-adjusted discretization, in which Euler--Maruyama proposals in Dikin geometry are corrected to remove time-step bias and recover the exact invariant law. We then show how the Dikin--Langevin diffusion can be used in an annealed interacting ensemble method for nonconvex optimization, where periodic resampling promotes particles that have entered favorable basins while retaining stochastic exploration. Implementations for all algorithms and scripts to reproduce the numerical experiments are available in the accompanying GitHub repository.\footnote{Code and experiment scripts are available at \url{https://github.com/infamoussoap/ConstrainedLangevin}}}

\section{Background}
In unconstrained $\mathbb{R}^d$, a common strategy for minimizing $f(x)$ is gradient descent,
\begin{equation*}
    x_{t+1}\ =\ x_t\ -\ \gamma\nabla f(x_t),
\end{equation*}
for $\gamma>0$, where the latter is the learning rate. This converges to local minima and global minima for convex $f(x)$, under mild conditions on $f$ \cite{Nesterov2004}. However, for nonconvex $f$, this purely deterministic method can become trapped in suboptimal basins. Stochastic approaches address this limitation by injecting randomness, enabling the algorithm to explore beyond local minima. 

For the remainder of the paper, unless otherwise stated, we restrict attention to $\rho(x)=\rho_1(x)=\exp(-f(x))/Z_1$.

\subsection{Metropolis--Hastings}
The Metropolis--Hastings \cite{Hastings1970, Metropolis1953} algorithm generates a discrete Markov chain $\{X_i\}_{i\in\mathbb{N}}$ on $U$ that has $\rho(x)$ as its stationary distribution, \textit{i.e.}, $X_i\to \rho$ in distribution as $i\to\infty$. Given the current state, $X_i$, one proposes a candidate $Y_i$, drawn from a proposal distribution $Q(\cdot \mid X_i)$. The proposal $Y_i$ is then accepted with probability
\begin{equation*}
A(X_i, Y_i)\ =\ \min\left(1, \frac{\rho(Y_i)}{\rho(X_i)}\frac{Q(X_i \mid Y_i)}{Q(Y_i \mid X_i)}\right),
\end{equation*}
and rejected otherwise. If accepted, the chain moves to $X_{i+1}=Y_i$; if rejected, it stays at $X_{i+1}=X_{i}$. This generic procedure is guaranteed to preserve $\rho$ as the invariant distribution of the Markov chain. 

A key flexibility lies in the choice of the proposal kernel $Q$, where improper choices can cause the chain to become stuck near the boundary of $U$. For example, taking $Q(\cdot|X_i)$ to be a Gaussian centered at $X_i$ (with fixed covariance) performs poorly near the boundary, since many proposals lie outside $U$ and are rejected, causing the chain to mix slowly.

\textit{Dikin Random Walks} \cite{Kannan2009, gu2024, jiang2024} modifies the proposal kernel $Q$ by using a Gaussian distribution whose covariance depends on the current state $X_i$. Specifically, $Q(\cdot|X_i)=\mathcal{N}(X_i,C(X_i))$, where 
\begin{equation}\label{eq:metropolis_proposal_and_barrier_hessian}
    C(x)=[\nabla^2 J(x)]^{-1},\quad\text{and}\quad J(x)\ =\ -\sum_{i=1}^K\log(b_i - a_i \cdot x).
\end{equation}
{Here, $J(x)$ is the logarithmic barrier function for $U$, and its Hessian $H(x)=\nabla^2J(x)$ induces a local geometry through the Dikin ellipsoid $E_x=\{y:(y-x)^TH(x)(y-x)\leq 1\}$. In the interior of $U$, this ellipsoid is relatively large and close to spherical; near the boundary, it contracts and becomes anisotropic, shrinking most strongly in directions pointing towards nearby constraints. As a result, the proposal is automatically scaled to the local geometry of $U$, reducing the chance of stepping outside the feasible region and improving exploration.}

\subsection{Langevin Dynamics}
The (overdamped) Langevin equation is given by the SDE
\begin{equation}\label{eq:langevin_sde}
    dX_t\ =\ -\nabla f(X_t)\, dt\ +\ \sqrt{2}\, dW_t,
\end{equation}
where $f:\mathbb{R}^d\to\mathbb{R}$ is the potential function (defined over all $\mathbb{R}^d$), and $W_t$ is standard Brownian motion on $\mathbb{R}^d$. Under mild regularity conditions on $f$, the random variable $X_t$ converges to the \emph{unconstrained} density proportional to $\exp(-f(x))$ as $t\to\infty$. In this regime, however, there exists $f$ (e.g., $f$ is strictly convex) where the particle escapes $U$ in finite time almost surely, even if the minimizer of $f$ lies in $U$. Therefore, to robustly handle constrained domains, the Langevin process must be modified either by reflecting at the boundary, altering the drift or diffusion, or redefining the geometry.

\textit{Reflected boundary conditions} \cite{Lamperski2020, Leimkuhler2023} for Eqn.~\eqref{eq:langevin_sde} enforce the constraint while preserving $\rho(x)$ as the stationary distribution. The simulation requires detecting boundary-hitting times, computing reflection directions for curved or polyhedral faces, and using discretization schemes that avoid overshooting during reflections. High-dimensional settings amplify these costs because boundary-intersection tests become computationally expensive. The reflection mechanism couples position and noise, preventing a clear formulation of a Metropolis--Hastings correction without breaking detailed balance. Reflective Langevin algorithms therefore suffer from bias and reduced sampling efficiency, particularly near boundaries due to frequent reflections.

\textit{Relaxation methods} \cite{gurbuzbalaban2024} incorporate the log-barrier function $J(x)$ into the potential and evolve the SDE
\begin{equation*}
dX_t = -\nabla H_{\lambda}(X_t)\, dt + \sqrt{2}\, dW_t,  
\quad\text{where}\quad H_\lambda(x) = f(x) - \lambda\, J(x),    
\end{equation*}
with $\lambda > 0$. For all $\lambda > 0$, the barrier term $J(x)$ diverges near the boundary, generating a strong inward drift that confines the continuous process to $U$. Large $\lambda$ values cause the stationary distribution to be dominated by $J(x)$, so the SDE neglects the objective $f(x)$. Small $\lambda$ values produce a barrier that is too weak to counteract the diffusive term in a discretized simulation, causing numerical trajectories to cross the boundary even though the continuous SDE would remain inside. Parameter tuning of $\lambda$ is therefore needed to ensure that the discretized dynamics both enforce the constraint and remain focused on minimizing $f(x)$.

{
\textit{Preconditioned Langevin dynamics} replaces the standard Langevin diffusion by
\begin{equation*}
    dX_t\ =\ -C\,\nabla f(X_t)\, dt\ +\ \sqrt{2C}\,dW_t,
\end{equation*}
where $C\in S^{d\times d}_{+}$ is a fixed symmetric positive definite matrix, and $X_0\in\mathbb{R}^d$. Under mild assumptions on $f$, the diffusion has an invariant \emph{unconstrained} density proportional to $\exp(-f(x))$. The matrix $C$ acts as a preconditioner, modifying the geometry of the dynamics, and improves convergence to equilibrium \cite{chok2025, lelievre2024optimizing}. 

More relevant to this paper is the case in which the preconditioner is position dependent, \textit{i.e.}, $C=C(X_t)$. In this setting, preserving the \emph{unconstrained} target density $\exp(-f(x))$ requires an additional drift correction, leading to the SDE \cite{Xifara2014, Roy2022, Girolami2011} 
\begin{align}\label{eq:alternative_riemannian_langevin}
    \begin{split}
    dX_t &= -C(X_t)\nabla f(X_t)dt + \Theta(X_t)dt + \sqrt{2C(X_t)}dW_t,\quad \Theta_i(x) = \sum_j\frac{\partial}{\partial X_j}C_{i,j}(X_t),
    \end{split}
\end{align}
equivalently, $\Theta(x)=\nabla\cdot C(x)$. In particular, \cite{Roy2022, Girolami2011} show that when $C:\mathbb{R}^d\to S^{d\times d}_{+}$, the resulting diffusion has invariant \emph{unconstrained} density proportional to $\exp(-f(x))$.

Existing literature primarily uses such position-dependent preconditioners to accelerate Langevin sampling. By contrast, this paper uses positive semidefinite choices of $C(X_t)$ to constrain the dynamics to compact domains. }

\section{Proposed Method}\label{sec:proposed_method}
\subsection{Overdamped Langevin Dynamics}

We now specialize \eqref{eq:alternative_riemannian_langevin} to the case where the preconditioner is given by the inverse Hessian of the logarithmic barrier. We call the resulting diffusion the \emph{Dikin--Langevin SDE}:
\begin{equation}\label{eq:dikin_langevin_sde}
    dX_t
    =
    -C(X_t)\nabla f(X_t)\,dt
    + \nabla\!\cdot C(X_t)\,dt
    + \sqrt{2\,C(X_t)}\,dW_t,
    \qquad 0\le t<T,
\end{equation}
with initial condition $X_0\in U^\circ:=U\setminus\partial U$, where $T:=\inf\{t\ge 0:\operatorname{dist}(X_t,\partial U)=0\}$ is the first boundary hitting time. {Here $C(x)$ is defined in \eqref{eq:metropolis_proposal_and_barrier_hessian}. In particular, if
\begin{equation*}
    J(x)=-\sum_{i=1}^K \log s_i(x), \qquad s_i(x):=b_i-a_i^\top x,
\end{equation*}
then
\begin{equation*}
    C(x)^{-1} = \nabla^2 J(x) = \sum_{i=1}^K \frac{a_i a_i^\top}{s_i(x)^2},\quad\text{and} \quad \nabla\!\cdot C(x) = -2\,C(x)\sum_{i=1}^K \left(\frac{a_i^\top C(x)a_i}{s_i(x)^3}\right)a_i.
\end{equation*}
}

Since the coefficients are smooth and locally Lipschitz on $U^\circ$, standard existence and uniqueness theory yields a unique continuous strong solution on $[0,T)$; see, for example, Theorem 3.1 of \cite{Watanabe2011} or Theorem 2.5 of \cite{Karatzas2004}.

{The key feature of \eqref{eq:dikin_langevin_sde} is that the barrier geometry degenerates in the normal direction as the boundary is approached. This degeneracy is essential: rather than being a numerical inconvenience, it is precisely what confines the dynamics to the feasible region. The following theorem shows that trajectories started in the interior do not hit the boundary in finite time almost surely.}

\begin{theorem}\label{theorem:invariant_subspace}
    Let $X_0\in U^\circ$. Then the solution to \eqref{eq:dikin_langevin_sde} satisfies
    \begin{equation}
        \mathbb{P}\!\left(X_t\in U^\circ \text{ for all } t\ge 0\right)=1.
    \end{equation}
    Equivalently, $T=\infty$ almost surely.
\end{theorem}

{
The proof is given in Appendix~\ref{sec:proof_theorem_invariant_subspace}. Heuristically, near a face with slack $s_i(x)=b_i-a_i^\top x$, both the drift and diffusion in the normal direction are of order $s_i(x)$. Thus, the normal motion slows as the process approaches the boundary, but it never reaches the boundary in finite time. In this sense, the boundary is inaccessible rather than reflecting or absorbing. Formally, this implies that the diffusion is a feasible constrained process with invariant density proportional to $e^{-f(x)}\mathbbm{1}_U(x)$.}

\begin{remark}
More generally, to target the tempered distribution $\rho_\beta(x)\propto e^{-\beta f(x)}\mathbbm{1}_U(x)$, one may consider either
\begin{equation}\label{eq:dikin_langevin_sde_temperature}
    dX_t = -C(X_t)\nabla f(X_t)\,dt + \frac{1}{\beta}\,\nabla\!\cdot C(X_t)\,dt + \sqrt{\frac{2\,C(X_t)}{\beta}}\,dW_t,
\end{equation}
or equivalently
\begin{equation}\label{eq:dikin_langevin_sde_temperature_v2}
    dX_t = -\beta\,C(X_t)\nabla f(X_t)\,dt + \nabla\!\cdot C(X_t)\,dt + \sqrt{2\,C(X_t)}\,dW_t.
\end{equation}
Both SDEs preserve $\rho_\beta(x)\mathbbm{1}_U(x)$ as invariant density. In practice, however, \eqref{eq:dikin_langevin_sde_temperature} is preferable, since the alternative scaling in \eqref{eq:dikin_langevin_sde_temperature_v2} amplifies the drift term as $\beta\to\infty$, which can make discretization less stable.
\end{remark}

\subsection{Metropolis-adjusted Dikin--Langevin}\label{sec:metropolis_dikin_langevin}
{
An Euler--Maruyama discretization of the Dikin--Langevin SDE does not, in general, preserve the target distribution exactly; instead, it introduces a time-step bias. To remove this bias, we apply a Metropolis--Hastings correction to the proposal. Concretely, given the current state $X_k$, we propose
\begin{equation}
    Y_k = X_k - h\!\left( C_\varepsilon(X_k)\nabla f(X_k) - \nabla\!\cdot C_\varepsilon(X_k) \right) + \sqrt{2h}\,C_\varepsilon(X_k)^{1/2}\xi_k,
\end{equation}
where $\xi_k \sim \mathcal N(0,I_d)$, $I_d$ is the $d\times d$ identity matrix, and
\begin{equation*}
    C_\varepsilon(x)^{-1} = \nabla^2 J(x)+\varepsilon I_d
\end{equation*}
is the regularized inverse barrier Hessian. Equivalently, conditional on $X_k=x$, the proposal is Gaussian,
\begin{equation*}
    Y_k \mid X_k=x \sim \mathcal N\!\bigl(\mu_h(x),\,2h\,C_\varepsilon(x)\bigr),
\end{equation*}
with mean
\begin{equation*}
    \mu_h(x) = x - h\!\left(C_\varepsilon(x)\nabla f(x) - \nabla\!\cdot C_\varepsilon(x) \right).
\end{equation*}
}

{
Let $q_h(\cdot\mid x)$ denote the density of $\mathcal N(\mu_h(x),2hC_\varepsilon(x))$. If $Y_k \notin U$, the proposal is rejected immediately. Otherwise, it is accepted with probability
\begin{equation*}\label{eq:accept_mdl}
    \alpha(X_k,Y_k)
    =
    \min\!\left(
        1,\,
        \frac{\rho(Y_k)\,q_h(X_k\mid Y_k)}
             {\rho(X_k)\,q_h(Y_k\mid X_k)}
    \right).
\end{equation*}
Hence, the resulting Markov chain has $\rho$ as its invariant distribution. A summary can be found in Algorithm~\ref{alg:metropolis_hasting_dikin_langevin}.

\begin{remark}
Although the Metropolis-adjusted Dikin--Langevin proposal is constructed from the Dikin geometry, it does not in general reduce to the Dikin random walk, even when $f$ is constant. Indeed, the proposal mean still contains the divergence term $\nabla\!\cdot C_\varepsilon(x)$, so the chain retains a nontrivial drift induced by the state-dependent geometry. Thus, the method should be viewed as a Metropolis-adjusted Langevin scheme in Dikin geometry, rather than as a Dikin random walk.
\end{remark}
}

Since $C_\varepsilon(x)$ is positive definite for every $x\in U^\circ$, we have
$
q_h(y\mid x)>0
$
for all $x,y\in U^\circ$. It follows that the Metropolis--Hastings kernel is $\rho$-irreducible on $U^\circ$. Moreover, because rejected proposals produce self-transitions, the chain is aperiodic. Under these standard conditions, $\rho$ is the unique stationary distribution; see, for example, Theorem 4 of \cite{Roberts_Rosenthal_2004}.

\begin{algorithm}
\caption{Metropolis--Hastings Dikin--Langevin sampler}
\label{alg:metropolis_hasting_dikin_langevin}
\begin{algorithmic}[1]
\Require Maximum step size $h_{\max}>0$, number of iterations $N$, initial point $x_0\in \mathrm{int}(U)\subset\mathbb{R}^d$, inverse-metric regularization $\varepsilon>0$, objective $f:\mathrm{int}(U)\to\mathbb{R}$, polytope inequalities $A\in\mathbb{R}^{k\times d}$, $b\in\mathbb{R}^d$

\Ensure Samples $\{x_k\}_{k=1}^N$

\Statex
\State Set $x\gets x_0$
\For{$k=1,\dots,N$}
    \State Compute $(C_{\varepsilon}(x)^{-1},L(x),m(x),f(x))\gets \Call{ProposalState}{x}$
    \State Sample $h_k\sim \operatorname{Unif}(0,h_{\max})$
    \State $y\gets \Call{Propose}{x,h_k,C_{\varepsilon}(x)^{-1},L(x),m(x)}$

    \If{$y\notin  U$}
        \State $x_k\gets x$ \Comment{Reject Proposal}
        \State \textbf{continue}
    \EndIf

    \State Compute $(C_{\varepsilon}(y)^{-1},L(y),m(y),f(y))\gets \Call{ProposalState}{y}$

    \State Compute $q_h(y\mid x)$, which is the density $\mathcal{N}(x+hm(x),2hC_{\varepsilon}(x))$

    \State Compute the acceptance probability
    \[
        \alpha(x,y)=
        \min\!\left\{
        1,\,
        \frac{e^{-f(y)}\,q_{h_k}(x\mid y)}
             {e^{-f(x)}\,q_{h_k}(y\mid x)}
        \right\}
    \]
    \State Sample $u\sim \operatorname{Unif}(0,1)$
    \If{$u<\alpha(x,y)$}
        \State $x\gets y$
    \EndIf

    \State $x_k\gets x$
\EndFor

\State \Return $\{x_k\}_{k=1}^N$

\Statex
\Function{ProposalState}{$x$}
    \State $s(x)\gets b-Ax$
    \State $C_{\varepsilon}(x)^{-1}\gets A^\top \operatorname{diag}(s(x)^{-2})A+\varepsilon I_d$
    \State Compute the Cholesky factorization: $C_{\varepsilon}(x)^{-1}=L(x)L(x)^\top$
    \State Compute $C_{\varepsilon}(x)\nabla f(x)$ and $\nabla\!\cdot C_{\varepsilon}(x)$ using Cholesky solves
    \State $m(x)\gets -C_{\varepsilon}(x)\nabla f(x)+\,\nabla\!\cdot C_{\varepsilon}(x)$
    \State \Return $\bigl(C_{\varepsilon}(x)^{-1},L(x),m(x),\log\pi(x)\bigr)$
\EndFunction

\Statex
\Function{Propose}{$x,h,C_{\varepsilon}(x)^{-1},L(x),m(x)$}
    \State Sample $\xi\sim\mathcal N(0,I_d)$
    \State Compute $\eta$ from the triangular solve: $L(x)^\top \eta=\xi$
    \Comment{so $\eta\sim\mathcal N(0,C_{\varepsilon}(x))$}
    \State \Return $y=x+h\,m(x)+\sqrt{2 h}\,\eta$
\EndFunction

\end{algorithmic}
\end{algorithm}

\paragraph{Randomized step size.}
In practice, using a fixed step size can lead to poor short-run exploration: the chain may remain nearly immobile for long periods and then move only in short bursts. Following \cite{chok2025}, we therefore randomize the step size by drawing
\[
    h_k \sim \mathrm{Unif}(0,h_{\max}),
\]
independently at each iteration, so that $\mathbb E[h_k]=h_{\max}/2$. This produces a mixture of Metropolis--Hastings kernels, each of which preserves $\rho$, and therefore the resulting randomized-step chain also has $\rho$ as its invariant distribution. Empirically, this randomization reduces resonance between the local proposal geometry and the discretization scale, leading to more robust exploration.

{
\begin{remark}[Implementation]
For each proposal, the main computational tasks are:
\begin{itemize}
    \item evaluating $f(X_k)$ and $\nabla f(X_k)$;
    \item forming $C_\varepsilon(X_k)=(\nabla^2J(X_k)+\varepsilon I_d)^{-1}$;
    \item sampling the Gaussian increment $C_\varepsilon(X_k)^{1/2}\xi_k$;
    \item evaluating $\det(C_\varepsilon(X_k))$, which appears in the proposal density.
\end{itemize}

Rather than forming $C_\varepsilon(x)$ explicitly, we compute the Cholesky factorization
\[
    \nabla^2J(x)+\varepsilon I_d = L(x)L(x)^\top.
\]
This factorization costs $O(d^3)$, with leading term $d^3/3$. Once $L(x)$ is available, the remaining linear-algebra operations are cheaper: $C_\varepsilon(x)\nabla f(x)$ and $\nabla\!\cdot C_\varepsilon(x)$ can be obtained by triangular solves in $O(d^2)$, the Gaussian increment can be generated via a triangular solve in $O(d^2)$, and
\[
    \log \det C_\varepsilon(x)
    =
    -2\sum_{i=1}^d \log L_{ii}(x).
\]
Thus, after forming $\nabla^2J(x)+\varepsilon I_d$, the per-step cost is dominated by the Cholesky factorization, while the remaining operations are lower-order.

As such, apart from the cost of evaluating $\nabla f$ (which we obtain using auto-differentiation), the adjusted Dikin--Langevin sampler has the same asymptotic computational scaling as the Dikin random walk.
\end{remark}}

\subsection{Stochastic Nonconvex Optimizer}
{
To treat nonconvex objectives over the constrained set $U$, we consider an interacting ensemble variant of the Dikin--Langevin method. Starting from a common initial condition $x_0 \in U^\circ$, we evolve $N$ particles in parallel. Between interaction times, each particle follows one Euler--Maruyama step of the annealed Dikin--Langevin dynamics,
\begin{equation}
    \widetilde{X}_{k+1}^{(n)}=X_k^{(n)}-h\!\left(C(X_k^{(n)})\nabla F(X_k^{(n)}) -\frac{1}{\beta_k}\nabla\!\cdot C(X_k^{(n)})\right) +\sqrt{\frac{2h\,C(X_k^{(n)})}{\beta_k}}\,\xi_k^{(n)}
\end{equation}
where $\{1/\beta_k\}_{k\in\mathbb N}$ is a temperature sequence decreasing monotonically to zero \cite{Hajek1988}, and
$\xi_k^{(n)} \sim \mathcal N(0, I)$.
If the proposal $\widetilde X_{k+1}^{(n)}$ lies in $U$, we set $X_{k+1}^{(n)}=\widetilde X_{k+1}^{(n)}$; otherwise, we reject the proposal and keep the previous state, $X_{k+1}^{(n)}=X_k^{(n)}$.

In the zero-temperature limit $\beta_k \to \infty$, this reduces to the deterministic barrier-preconditioned descent dynamics for constrained optimization \cite{Alvarez2004}. For $\beta_k<\infty$, the noise term allows the particles to escape shallow or suboptimal local minima, while the decreasing temperature gradually shifts the dynamics from exploration toward exploitation.

To couple exploration with selection, we introduce an interaction step every $R$ iterations. At such times, we assign weights
\[
    w_k^{(n)} = \frac{\exp\!\bigl(-f(X_k^{(n)})\bigr)}{\sum_{m=1}^N\exp\!\bigl(-f(X_k^{(m)})\bigr)},
\]
and resample the ensemble accordingly. The particle with the largest weight is retained deterministically, while the remaining $N-1$ particles are drawn with replacement from the current ensemble using $\{w_k^{(n)}\}_{n-1}^N$. Hence, particles that have reached lower objective regions are replicated, whereas less favorable particles are discarded. The resampled ensemble then provides the initial condition for the next block of Euler--Maruyama updates. A summary can be found in Algorithm ~\ref{alg:interacting_dikin_langevin}.

This interaction mechanism balances two effects: the stochastic Dikin--Langevin evolution promotes exploration of the feasible region, while the periodic resampling concentrates computational effort on particles that have already moved into promising basins. In this sense, the method combines an annealed barrier-based diffusion with an elitist selection step, drawing inspiration from both mean-field optimization methods \cite{Fornasier2021, Huang2022, Borghi2023, Ko2025} and genetic algorithms \cite{BHANDARI1996, Rudolph1994, Syswerda1991}.}

\begin{algorithm}
\caption{Unadjusted Interacting Dikin--Langevin Algorithm}
\label{alg:interacting_dikin_langevin}
\begin{algorithmic}[1]
\Require Step size $h>0$, number of iterations $K$, number of chains $N$, initial point $x_0\in \mathrm{int}(U)\subset\mathbb{R}^d$, inverse-metric regularization $\varepsilon>0$, resampling period $R\in\mathbb{N}$, temperature sequence $(1/\beta_k)_{k=0}^{K-1}$, objective $f:\mathrm{int}(U)\to\mathbb{R}$, polytope inequalities $A\in\mathbb{R}^{k\times d}$, $b\in\mathbb{R}^d$
\State Initialize all chains at the same point: $X_0^{(n)} \gets x_0,\quad n=1,\dots,N$
\For{$k=0,1,\dots,K-1$}
    \For{$n=1,\dots,N$}
        \State Sample noise: $\xi_k^{(n)} \sim \mathcal{N}(0,I_d)$.
        \State Euler--Maruyama: $$\widetilde X_{k+1}^{(n)}=X_k^{(n)}-h\!\left(C(X_k^{(n)})\nabla f(X_k^{(n)}) -\frac{1}{\beta_k}\nabla\!\cdot C(X_k^{(n)})\right) +\sqrt{\frac{2hC(X_k^{(n)})}{\beta_k}}\,\xi_k^{(n)}.$$
        \If{$\widetilde X_{k+1}^{(n)}\in U$}
            \State $X_{k+1}^{(n)} \gets \widetilde X_{k+1}^{(n)}$
        \Else
            \State $X_{k+1}^{(n)} \gets X_k^{(n)}$
        \EndIf
    \EndFor
    \If{$(k+1)\bmod R = 0$}
        \State Compute weights: $\widehat{w}_n = \exp\!\bigl(-f(X_{k+1}^{(n)})\bigr),\qquad n=1,\dots,N.$
        \State Normalize: $w_n=\widehat{w}_n/\sum_{m=1}^N\widehat{w}_m$.
        \State Let $n^\star \in \arg\max_{1\le n\le N}\ w_n.$
        \State Keep the best chain: $Y_{k+1}^{(1)} \gets X_{k+1}^{(n^\star)}.$
        \For{$j=2,\dots,N$}
            \State Sample an index $I_j\in\{1,\dots,N\}$ with $\mathbb{P}(I_j=n)=w_n.$
            \State Set $Y_{k+1}^{(j)} \gets X_{k+1}^{(I_j)}.$
        \EndFor
        \State Replace the ensemble: $X_{k+1}^{(n)} \gets Y_{k+1}^{(n)},\qquad n=1,\dots,N.$
    \EndIf
\EndFor
\State \Return $\{X_k^{(n)}: k=0,\dots,K,\; n=1,\dots,N\}$
\end{algorithmic}
\end{algorithm}

\section{Applications}
\subsection{Unit Ball Constraint}

As a preliminary numerical experiment, we illustrate the behavior of the unadjusted Dikin--Langevin SDE on a simple constrained target: the standard Gaussian distribution in $d=20$, truncated to the unit ball $B_1(0)$. This example serves as a benchmark for continuous-time diffusion and, in particular, for the discretization bias introduced by Euler--Maruyama time-stepping.

By radial symmetry, the normalizing constant is
\begin{equation*}
    Z_1\ =\ \int_{B_1(0)}\exp(-\|x\|^2/2)\, dx\ =\ \frac{2\pi^{d/2}}{\Gamma(d/2)}\int_{0}^1r^{d-1}\exp(-r^2/2)\, dr\ =\ \frac{(2 \pi )^{d/2} \left(\Gamma \left(\frac{d}{2}\right)-\Gamma \left(\frac{d}{2},\frac{1}{2}\right)\right)}{\Gamma \left(\frac{d}{2}\right)}.
\end{equation*}
and the corresponding benchmark expectation is
\begin{equation*}
    \mathbb{E}[\|x\|]\ =\ \frac{1}{Z_1}\int_{B_1(0)}\|x\|\exp(-\|x\|^2/2)\, dx\ =\ \frac{\sqrt{2} \left(\Gamma \left(\frac{d+1}{2}\right)-\Gamma \left(\frac{d+1}{2},\frac{1}{2}\right)\right)}{\Gamma \left(\frac{d}{2}\right)-\Gamma \left(\frac{d}{2},\frac{1}{2}\right)}.
\end{equation*}

{
We simulate the Dikin--Langevin SDE using the barrier $J(x)=-\log(1-\|x\|^2)$
with
\begin{equation*}
    C(x)=\frac{1-\|x\|^2}{2}\left(I-\frac{2}{1+\|x\|}xx^\top\right),\quad\text{and}\quad \nabla\cdot C(x)=-\frac{(1-\|x\|^2)\Big(d(1+\|x\|^2)+2\Big)}{(1+\|x\|^2)^2}x,
\end{equation*}
together with an Euler--Maruyama discretization. Since the unadjusted discretization does not preserve the exact invariant law, one expects a step-size-dependent asymptotic bias. To quantify this effect, we run the scheme with several time steps up to the final time $T=5000$, recording one sample every $\Delta t=0.1$ units of integration time.}

\begin{figure}
    \centering
    \includegraphics[width=0.7\linewidth]{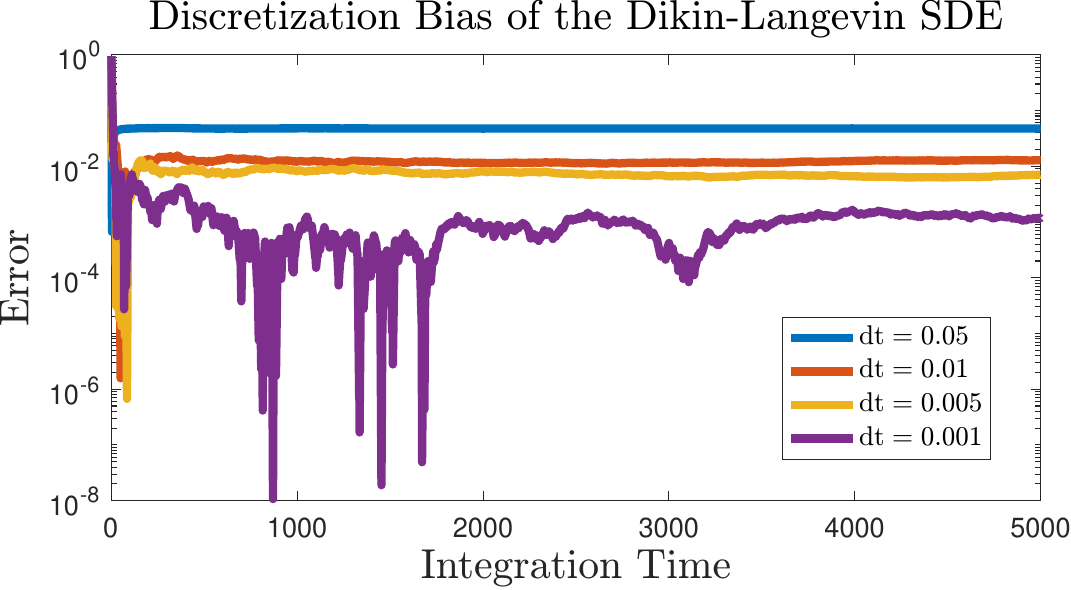}
    \caption{Discretization bias of the unadjusted Euler--Maruyama approximation of the Dikin--Langevin SDE for several time steps $\mathrm{dt}$, when sampling from a $20$-dimensional standard Gaussian truncated to the unit ball. The error is $\Big{|}{E}[\|x\|]-\frac{1}{N}\sum_{n=1}^N\|X_{n}\|\Big{|}$, where $X_n$ is the sample at integration time $t=0.1n$. Smaller time steps produce smaller asymptotic bias, consistent with first-order discretization error.}
    \label{fig:Gaussian_samples}
\end{figure}

Figure~\ref{fig:Gaussian_samples} shows the absolute error in the running estimator of $\mathbb E[\|X\|]$. In each case, the estimator approaches the correct benchmark value, but a nonzero asymptotic bias remains, as expected for a first-order discretization. Moreover, the bias increases with the time step. In particular, reducing the step size from $0.01$ to $0.001$ decreases the final error from $1.28\times 10^{-2}$ to $1.18\times 10^{-3}$, which is consistent with first-order behavior in the discretization parameter.

\subsection{Metropolis-Adjusted Samplers}

We now compare three Metropolis-adjusted samplers on a highly anisotropic box-constrained target. The feasible set is
\begin{equation*}
    U=\left\{x\in\mathbb{R}^{10}:\,-b_i\le x_i\le b_i,\quad i=1,\dots,10\right\},
\end{equation*}
where $\{b_i\}_{i=1}^{10}$ are logarithmically spaced between $b_1=1$ and $b_{10}=0.01$. This creates several severely constrained coordinates, with the smallest side lengths serving as the main bottleneck for mixing. On this domain, we consider the Gaussian target
\begin{equation*}
    \rho(x)\propto \exp\!\left(-\sum_{i=1}^{10}\frac{(x_i-\mu_i)^2}{2\sigma_i^2}\right),\quad\text{with}\quad \mu_i=0.5\,b_i,
    \qquad
    \sigma_i=0.5\,b_i^{3/2}.
\end{equation*}
Thus, the target is centered away from the origin but becomes increasingly concentrated in the most tightly constrained coordinates.

\begin{figure}
    \centering
    \includegraphics[width=0.95\linewidth]{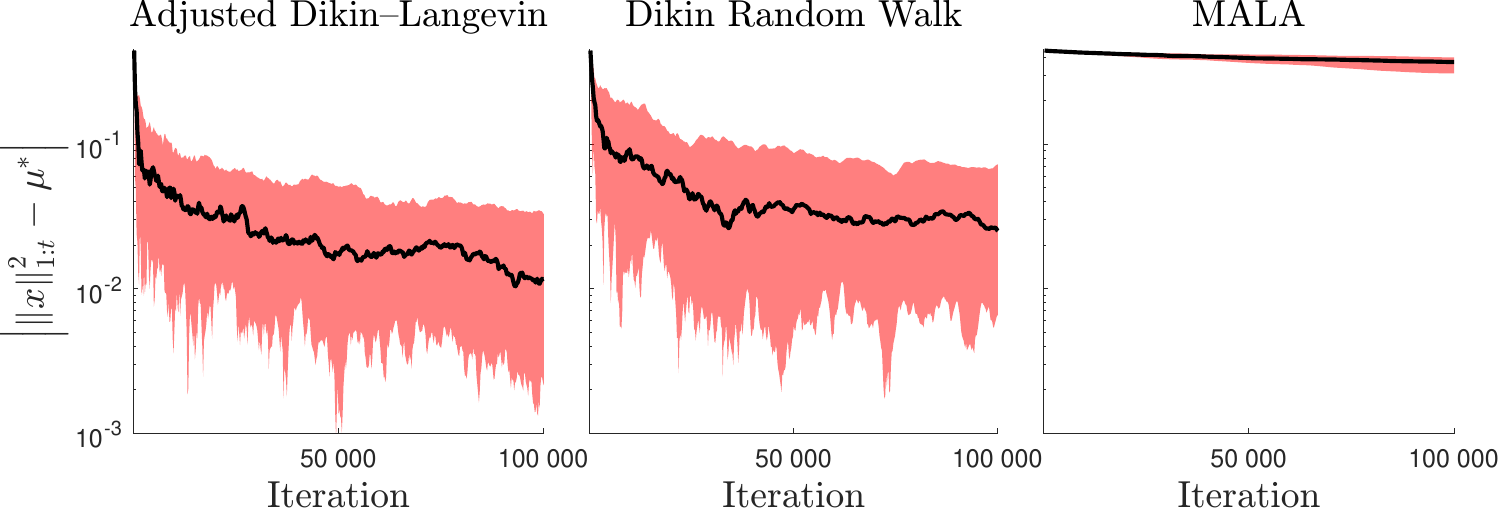}
    \caption{Convergence of the rolling-mean estimator on a log scale. For each algorithm, the black curve is the median of $| {\|x\|}^2_{1:t}-\mu^\ast |$ over $50$ independent runs, where ${\|x\|}_{1:t}^2=t^{-1}\sum_{i=1}^t \|x_i\|^2$ and $\mu^\ast$ is the ground-truth expectation. The red band marks the interdecile range (10th-90th percentiles).}
    \label{fig:metropolis_results}
\end{figure}

{
To assess convergence across the independent runs, we consider two diagnostics: (1) the error in the running estimator
\begin{equation}
    \bar m_t=\frac{1}{t}\sum_{i=1}^t \|x_i\|^2,
\end{equation}
of $\mathbb{E}[\|x\|^2]$, and (2) the rank-normalized split-$\hat R$ values \cite{Vehtari2021}. This statistic compares within-chain and between-chain variability; values close to $1$ indicate that the chains have largely forgotten their initializations and are exploring the same distribution.

We compare the adjusted Dikin--Langevin (ADL) sampler from Section~\ref{sec:metropolis_dikin_langevin} against the Dikin random walk (DRW) and against standard MALA \cite{besag1994mala,parisi1981ula}. For the barrier-based methods, we fix $\varepsilon=10^{-5}$. The proposal maximum randomized step-size, $h_{\max}$, for each sampler is tuned to achieve an average acceptance rate close to $0.6$. Each method is initialized at the origin and runs for $100{,}000$ iterations. To quantify variability, each method is repeated across 50 independent runs, each seeded for reproducibility.

\textit{Convergence diagnostics.}
Table~\ref{tab:rhat_values} reports rank-normalized split-$\hat R$ values computed coordinatewise from the last $50{,}000$ draws. Among the three methods, ADL shows the most uniform convergence: its $\hat R$ values remain tightly clustered near $1$, with a median of $1.001$, a 90th percentile of $1.008$, and a maximum of $1.009$. DRW performs reasonably in the bulk, but its heavier upper tail indicates less reliable mixing in the most constrained directions. By contrast, MALA mixes poorly under the same acceptance-rate tuning, with very large upper-tail $\hat R$ values, showing that an isotropic proposal struggles to adapt to the geometry of the narrow coordinates.

\textit{Trajectory diagnostics.}
Figure~\ref{fig:metropolis_results} shows the error in the running estimator $\bar{m}_t$ relative to the ground-truth expectation $\mu^\ast\approx 0.44$. The trajectories are consistent with the $\hat R$ diagnostics. ADL exhibits the fastest and most stable decay of the estimation error, together with a narrowing interdecile band across runs. DRW improves more slowly and retains visibly greater dispersion at long times. MALA remains both inaccurate and highly variable throughout the run, reflecting its failure to adapt to the anisotropic box geometry. Overall, these results indicate that incorporating the barrier-based preconditioner into the Metropolis-adjusted proposal substantially improves robustness on strongly constrained targets.}

\begin{table}
    \centering
    \caption{Rank-normalized split-$\hat R$ computed per dimension on the last $50\,000$ draws across $50$ independent runs. We report the median, 90th percentile, and maximum value across all dimensions. Values $\leq 1.01$ indicate good mixing; larger values signal non-convergence.}
    \label{tab:rhat_values}
    \begin{tabular}{c|c|c|c|c}
        \toprule
Method & {median $\hat R$} & {90th percentile $\hat R$} & {max $\hat R$}  \\
\midrule
Adjusted Dikin--Langevin & 1.001 & 1.008 & 1.009 \\
Dikin Random Walk        & 1.006 & 1.032 & 1.036 \\
MALA                     & 1.061 & 3.656 & 3.663 \\
\bottomrule
    \end{tabular}
\end{table}

\subsection{Nonconvex Optimization}
\begin{figure}
    \centering
    \includegraphics[width=0.5\linewidth]{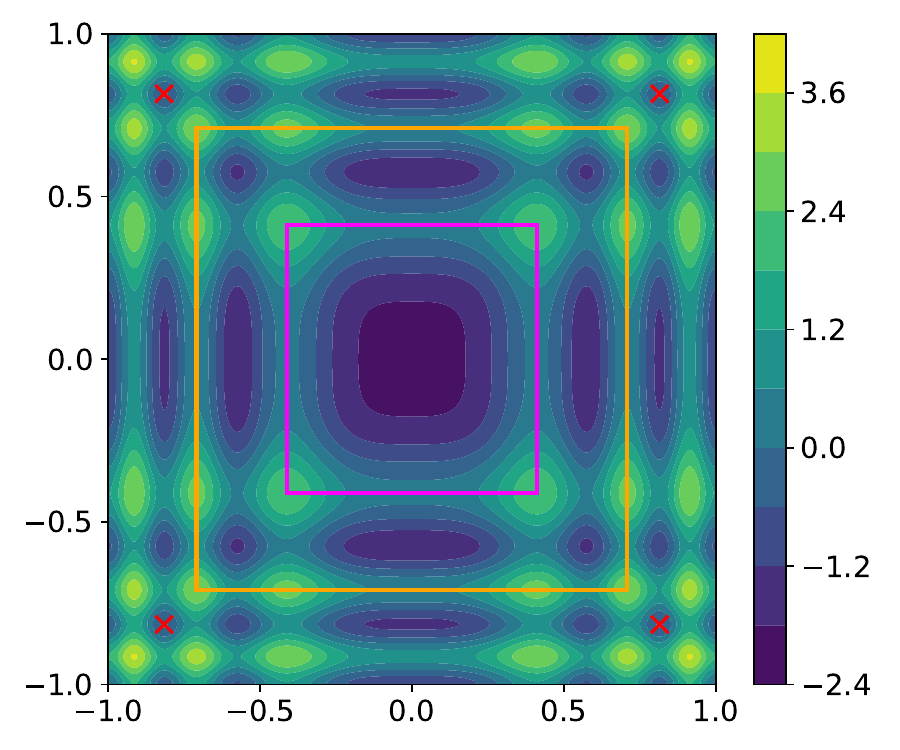}
    \caption{Two-dimensional analogue of the objective in \eqref{eq:nonconvex_optimization_problem}. The global minimizer is at $(0,0)$. The red crosses indicate the projected initialization points $(\pm 0.8156,\pm 0.8156)$, corresponding to starts near the suboptimal basin $x_i=\pm\sqrt{2/3}$. The orange square marks the outer escape region $|x_i|\le 1/\sqrt{2}$, and the magenta square marks the inner escape region $|x_i|\le 1/\sqrt{6}$.}
    \label{fig:multimodal_optimization}
\end{figure}

{
To illustrate how stochastic exploration helps the method escape suboptimal local minima, we consider the constrained nonconvex problem
\begin{equation}\label{eq:nonconvex_optimization_problem}
    \min_{x\in[-1,1]^{10}} f(x),
    \qquad
    f(x)=\sum_{i=1}^{10}\bigl(x_i^2-\cos(6\pi x_i^2)\bigr).
\end{equation}
Since $f$ is separable, its landscape is determined by the one-dimensional function
\[
    g(t)=t^2-\cos(6\pi t^2), \qquad t\in[-1,1].
\]
The global minimum occurs at $t=0$, and hence the unique global minimizer of \eqref{eq:nonconvex_optimization_problem} is $x=0$, with value $f(0)=-10$. In addition, $g$ has several non-global local minima, occurring at
\[
    t=\pm \frac{1}{\sqrt{3}},\quad
    t=\pm \sqrt{\frac{2}{3}},\quad
    t=\pm 1,
\]
as well as local maxima at
\[
    t=\pm \frac{1}{\sqrt{6}},\quad
    t=\pm \frac{1}{\sqrt{2}},\quad
    t=\pm \sqrt{\frac{5}{6}}.
\]
Thus \eqref{eq:nonconvex_optimization_problem} contains many suboptimal basins that can trap deterministic descent methods.

We compare the interacting and non-interacting Dikin--Langevin optimizers using $N=5$ chains. In each run, all five chains are initialized at the same point $X^0$, where each coordinate $X_i^0$ is chosen independently to be $0.8156$ or $-0.8156$ with equal probability. Hence, every run starts near the suboptimal local minimum $x_i=\pm\sqrt{2/3}$, and any performance gain must come from stochastic exploration and, in the interacting case, the resampling step rather than from favorable initialization.\footnote{We do not compare against the standard Langevin SDE, as we find that in practice its discretization typically leaves the feasible set.} To quantify variability, each algorithm is repeated over 100 independent runs, with each run seeded for reproducibility.

We use step size $h=0.01$, total iteration count $K=40\,000$, resampling period $R=1\,000$, and the annealing schedule
\begin{equation*}
    \frac{1}{\beta_k}=
    \begin{cases}
        \dfrac{2}{\log(3+2hk)}, & 0\le k<35\,000,\\[0.8em]
        \dfrac{2}{\log(3+2hk)^4}, & 35\,000\le k<39\,000,\\[0.8em]
        0, & 39\,000\le k\le 40\,000.
    \end{cases}
\end{equation*}
This schedule reflects the three phases of the algorithm. In the first phase, we use the classical logarithmic cooling rate $\beta_k\asymp \log(k)$ from simulated annealing for reflected Langevin dynamics \cite{Geman1986}. In the second phase, the temperature is reduced more aggressively to concentrate the particles within low-energy basins. In the final phase, we set $\beta_k=\infty$, so the noise vanishes, and the dynamics reduce to deterministic constrained gradient descent in the Dikin geometry \cite{Alvarez2004}, providing a final local refinement step.

For each run, we record three summary statistics:
\begin{enumerate}
    \item the best objective value attained across all five chains;
    \item the first-escape time from the outer local basin, defined as the first iteration at which at least one chain satisfies $|x_i|\le 1/\sqrt{2}$ for every $i$;
    \item the first-escape time from the inner local basin, defined as the first iteration at which at least one chain satisfies $|x_i|\le 1/\sqrt{6}$ for every $i$.
\end{enumerate}
These two basin thresholds are illustrated in Fig.~\ref{fig:multimodal_optimization}. Entering the inner basin places a chain in the neighborhood of the global minimizer, although, unlike deterministic gradient descent, the stochastic dynamics may still subsequently leave this region before cooling is complete.

Table~\ref{tab:optimization_result_values} shows that interaction primarily improves the \emph{reliability} of basin escape rather than the best-case performance. Both methods occasionally reach the global minimum and have the same best observed escape times, but the interacting method does so far more consistently: it reaches the global optimum in 87 of 100 runs, compared with only 36 of 100 runs without interaction. This improvement is already visible in the median best objective value, which is $-10.00$ with interaction versus $-9.98$ without interaction, and in the median second-escape time, which decreases from $1755$ to $1641$. The largest effect appears in the tail behavior of the first-escape time: with interaction, the worst run escapes the outer basin after $1942$ iterations, whereas this rises to $3827$ without interaction. Overall, the resampling step does not materially change the luckiest runs, but it substantially reduces the frequency with which the ensemble remains trapped in poor local basins.
}

\begin{table}
\centering
    \caption{Summary over 100 independent runs of the interacting and non-interacting Dikin--Langevin optimizers, each using five chains. For each run, we report the best objective value attained across the five chains, the first-escape time into the outer basin $|x_i|\le 1/\sqrt{2}$ for all $i$, and the first-escape time into the inner basin $|x_i|\le 1/\sqrt{6}$ for all $i$. The interacting method reaches the global optimum in 87 of 100 runs, compared with 36 of 100 runs for the non-interacting method.}
    \label{tab:optimization_result_values}
\begin{tabular}{l|ccc|ccc}
\toprule
 & \multicolumn{3}{c|}{Interaction} & \multicolumn{3}{c}{No Interaction} \\
 & Best Obj.& First Escape & Second Escape & Best Obj.& First Escape & Second Escape \\
\midrule
min & -10.00 & 35 & 149 & -10.00 & 35 & 149 \\
50\% & -10.00 & 189 & 1\,641 & -9.98 & 189 & 1\,755 \\
max & -9.97 & 1\,942 & 14\,872 & -9.81 & 3\,827 & 14\,623 \\
\bottomrule
\end{tabular}
\end{table}

\section{Conclusion}
We introduced a Dikin--Langevin approach to constrained stochastic computation on compact polyhedra, based on the inverse Hessian of the logarithmic barrier. In continuous time, this geometry yields a diffusion whose normal drift and diffusion both degenerate at the boundary in precisely the way needed to keep trajectories in the interior almost surely. This provides a clean alternative to reflected dynamics: feasibility is enforced by the geometry itself, rather than by boundary detection or projection.

{
For computation, we adopted a discretize-then-correct strategy. The resulting Metropolis-adjusted Dikin--Langevin sampler combines local barrier geometry with exact invariance of the target law, while remaining implementable using standard linear algebra primitives such as Cholesky factorizations and triangular solves. Numerically, this method outperforms both the Dikin random walk and standard MALA on anisotropic constrained Gaussian targets, indicating that the combination of local preconditioning and Metropolis correction is especially effective when some coordinates are much more tightly constrained than others.

We also showed that the same geometry can be used in an annealed interacting optimizer for nonconvex constrained problems. In that setting, stochastic exploration allows chains to leave poor local basins, while periodic resampling concentrates effort on promising regions. The numerical results indicate that interaction improves the reliability of global-basin discovery, rather than merely improving best-case behavior, which is precisely the regime of practical interest in multimodal optimization.

Several extensions remain natural. On the analytical side, it would be valuable to establish quantitative bounds on the mixing time of the Dikin--Langevin SDE and its Metropolis-adjusted version. The case $f=0$ is a particularly interesting benchmark, since the Metropolis-adjusted Dikin--Langevin sampler does not reduce to the Dikin random walk. Understanding this difference more precisely would clarify how much of the method's performance is attributable to the barrier geometry $C(x)$ alone and how much to its Langevin-type discretization. On the algorithmic side, underdamped Langevin and nonreversible MCMC variants are promising directions, as are scalable approximations of the barrier Hessian for problems with many faces. More broadly, the results here suggest that interior-point geometry offers a useful organizing principle for designing constrained stochastic algorithms that remain both geometrically faithful and computationally practical.
}

\section*{Acknowledgments}
The authors thank Dr. Amanda Lenzi for the helpful conversations.

\section*{Funding}
The authors declare no external funding.

\bibliographystyle{plain}

\begin{thebibliography}{10}

\bibitem{Adcock2025}
B.~Adcock, M.~J. Colbrook, and M.~Neyra-Nesterenko.
\newblock Restarts subject to approximate sharpness: A parameter-free and optimal scheme for first-order methods.
\newblock {\em Foundations of Computational Mathematics}, February 2025.

\bibitem{Alvarez2004}
F.~Alvarez, J.~Bolte, and O.~Brahic.
\newblock Hessian riemannian gradient flows in convex programming.
\newblock {\em SIAM Journal on Control and Optimization}, 43(2):477–501, January 2004.

\bibitem{Baravdish2023}
G.~Baravdish, B.~T. Johansson, O.~Svensson, and W.~Ssebunjo.
\newblock Identifying a response parameter in a model of brain tumour evolution under therapy.
\newblock {\em IMA Journal of Applied Mathematics}, 88(2):378–404, April 2023.

\bibitem{besag1994mala}
J.~Besag.
\newblock {``Comments on ``Representations of knowledge in complex systems" by U. Grenander and MI Miller}.
\newblock {\em Journal of the Royal Statistical Society, Series B.}, 56:591–592, 1994.

\bibitem{BHANDARI1996}
D.~Bhandari, C.~A. Murthy, and S.~K. Pal.
\newblock Genetic algorithm with elitist model and its convergence.
\newblock {\em International Journal of Pattern Recognition and Artificial Intelligence}, 10(06):731–747, September 1996.

\bibitem{BIGGS1978}
M.~C. Biggs.
\newblock On the convergence of some constrained minimization algorithms based on recursive quadratic programming.
\newblock {\em IMA Journal of Applied Mathematics}, 21(1):67–81, 1978.

\bibitem{Borghi2023}
G.~Borghi, M.~Herty, and L.~Pareschi.
\newblock Constrained consensus-based optimization.
\newblock {\em SIAM Journal on Optimization}, 33(1):211–236, January 2023.

\bibitem{BouRabee2009}
N.~Bou‐Rabee and E.~Vanden‐Eijnden.
\newblock Pathwise accuracy and ergodicity of {M}etropolized integrators for {SDE}s.
\newblock {\em Communications on Pure and Applied Mathematics}, 63(5):655–696, November 2009.

\bibitem{Chen2024}
P.-B. Chen, G.-H. Lin, W.~Xu, and X.~Zhu.
\newblock Supply chain network equilibrium with outsourcing for fresh agricultural products under stochastic demands.
\newblock {\em IMA Journal of Management Mathematics}, October 2024.

\bibitem{chok2025}
J.~Chok, M.~W. Lee, D.~Paulin, and G.~M. Vasil.
\newblock Divide, interact, sample: The two-system paradigm, 2025.

\bibitem{Cox2008}
D.~D. Cox, R.~M. Hardt, and P.~Klou\v{c}ek.
\newblock Convergence of {G}ibbs measures associated with simulated annealing.
\newblock {\em SIAM Journal on Mathematical Analysis}, 39(5):1472–1496, January 2008.

\bibitem{FLETCHER1971}
R.~Fletcher.
\newblock A general quadratic programming algorithm.
\newblock {\em IMA Journal of Applied Mathematics}, 7(1):76–91, 1971.

\bibitem{Fornasier2021}
M.~Fornasier, H.~Huang, L.~Pareschi, and P.~S\"{u}nnen.
\newblock Consensus-based optimization on the sphere: convergence to global minimizers and machine learning.
\newblock {\em J. Mach. Learn. Res.}, 22(1), January 2021.

\bibitem{Geman1986}
S.~Geman and C.-R. Hwang.
\newblock Diffusions for global optimization.
\newblock {\em SIAM Journal on Control and Optimization}, 24(5):1031–1043, July 1986.

\bibitem{Girolami2011}
M.~Girolami and B.~Calderhead.
\newblock Riemann manifold {L}angevin and {H}amiltonian {M}onte {C}arlo methods.
\newblock {\em Journal of the Royal Statistical Society Series B: Statistical Methodology}, 73(2):123–214, March 2011.

\bibitem{gu2024}
Y.~Gu, N.~L. Kuang, Y.-A. Ma, A.~Song, and L.~Zhang.
\newblock Log-concave sampling from a convex body with a barrier: a robust and unified {D}ikin walk, 2024.

\bibitem{gurbuzbalaban2024}
M.~G\"{u}rb\"{u}zbalaban, Y.~Hu, and L.~Zhu.
\newblock Penalized overdamped and underdamped {L}angevin {M}onte {C}arlo algorithms for constrained sampling.
\newblock {\em JMLR}, 25(1), January 2024.

\bibitem{Hajek1988}
B.~Hajek.
\newblock Cooling schedules for optimal annealing.
\newblock {\em Mathematics of Operations Research}, 13(2):311--329, 1988.

\bibitem{Hastings1970}
W.~K. Hastings.
\newblock {M}onte {C}arlo sampling methods using {M}arkov chains and their applications.
\newblock {\em Biometrika}, 57(1):97–109, April 1970.

\bibitem{Hintermuller2009}
M.~Hintermuller, V.~A. Kovtunenko, and K.~Kunisch.
\newblock A {P}apkovich-{N}euber-based numerical approach to cracks with contact in 3d.
\newblock {\em IMA Journal of Applied Mathematics}, 74(3):325–343, April 2009.

\bibitem{Holley1988}
R.~Holley and D.~Stroock.
\newblock Simulated annealing via {S}obolev inequalities.
\newblock {\em Communications in Mathematical Physics}, 115(4):553–569, December 1988.

\bibitem{Huang2022}
H.~Huang and J.~Qiu.
\newblock On the mean‐field limit for the consensus‐based optimization.
\newblock {\em Mathematical Methods in the Applied Sciences}, 45(12):7814–7831, April 2022.

\bibitem{jiang2024}
M.~Jiang and Y.~Chen.
\newblock Regularized {D}ikin walks for sampling truncated logconcave measures, mixed isoperimetry and beyond worst-case analysis, 2024.

\bibitem{Kannan2009}
R.~Kannan and H.~Narayanan.
\newblock Random walks on polytopes and an affine interior point method for linear programming.
\newblock In {\em Proceedings of the forty-first annual ACM symposium on Theory of computing}, STOC ’09, page 561–570. ACM, May 2009.

\bibitem{Karatzas2004}
I.~Karatzas and S.~E. Shreve.
\newblock {\em Brownian motion and stochastic calculus}.
\newblock Graduate texts in mathematics. Springer, New York, NY, 2 edition, August 2004.

\bibitem{Kirkpatrick1983}
S.~Kirkpatrick, C.~D. Gelatt, and M.~P. Vecchi.
\newblock Optimization by simulated annealing.
\newblock {\em Science}, 220(4598):671–680, May 1983.

\bibitem{Kovtunenko2006}
V.~A. Kovtunenko.
\newblock Primal–dual methods of shape sensitivity analysis for curvilinear cracks with nonpenetration.
\newblock {\em IMA Journal of Applied Mathematics}, 71(5):635–657, October 2006.

\bibitem{Ko2025}
M.~Koß, S.~Weissmann, and J.~Zech.
\newblock On the mean‐field limit of consensus‐based methods.
\newblock {\em Mathematical Methods in the Applied Sciences}, 49(5):4214–4240, November 2025.

\bibitem{Lamperski2020}
A.~G. Lamperski.
\newblock Projected stochastic gradient {L}angevin algorithms for constrained sampling and non-convex learning.
\newblock {\em CoRR}, abs/2012.12137, 2020.

\bibitem{Langevin1908}
P.~Langevin.
\newblock Sur la th\'eorie du mouvement brownien.
\newblock {\em C. R. Acad. Sci. (Paris) 146}, pages 530--533, 1908.

\bibitem{Leimkuhler2023}
B.~Leimkuhler, A.~Sharma, and M.~V. Tretyakov.
\newblock Simplest random walk for approximating robin boundary value problems and ergodic limits of reflected diffusions.
\newblock {\em The Annals of Applied Probability}, 33(3), June 2023.

\bibitem{lelievre2024optimizing}
T.~Leli\'evre, G.~A. Pavliotis, G.~Robin, R.~Santet, and G.~Stoltz.
\newblock Optimizing the diffusion coefficient of overdamped {L}angevin dynamics, 2024.

\bibitem{Lemons1997}
D.~S. Lemons and A.~Gythiel.
\newblock Paul {L}angevin’s 1908 paper “on the theory of {B}rownian motion” [“{S}ur la th\'eorie du mouvement brownien, ” c. r. acad. sci. (paris) 146, 530–533 (1908)].
\newblock {\em American Journal of Physics}, 65(11):1079–1081, November 1997.

\bibitem{Livingstone2021}
S.~Livingstone.
\newblock Geometric ergodicity of the random walk {M}etropolis with position-dependent proposal covariance.
\newblock {\em Mathematics}, 9(4):341, February 2021.

\bibitem{Metropolis1953}
N.~Metropolis, A.~W. Rosenbluth, M.~N. Rosenbluth, A.~H. Teller, and E.~Teller.
\newblock Equation of state calculations by fast computing machines.
\newblock {\em The Journal of Chemical Physics}, 21(6):1087–1092, June 1953.

\bibitem{Nesterov2004}
Y.~Nesterov.
\newblock {\em Introductory Lectures on Convex Optimization}.
\newblock Springer US, 2004.

\bibitem{ONeill2020}
M.~O'Neill and S.~J. Wright.
\newblock A log-barrier {N}ewton-{CG} method for bound constrained optimization with complexity guarantees.
\newblock {\em IMA Journal of Numerical Analysis}, 41(1):84–121, April 2020.

\bibitem{parisi1981ula}
G.~Parisi.
\newblock {Correlation Functions and Computer Simulations}.
\newblock {\em Nuclear Physics B}, 180:378, 1981.

\bibitem{Pavliotis2014}
G.~A. Pavliotis.
\newblock {\em Stochastic Processes and Applications: Diffusion Processes, the Fokker-Planck and Langevin Equations}.
\newblock Springer New York, 2014.

\bibitem{Robert2004}
C.~P. Robert and G.~Casella.
\newblock {\em Monte Carlo Statistical Methods}.
\newblock Springer New York, 2004.

\bibitem{Roberts_Rosenthal_2004}
G.~O. Roberts and J.~S. Rosenthal.
\newblock General state space {M}arkov chains and {MCMC} algorithms.
\newblock {\em Probability Surveys}, 1, January 2004.
\newblock arXiv:math/0404033.

\bibitem{roberts_1996_mala}
G.~O. Roberts and R.~L. Tweedie.
\newblock {Exponential convergence of {L}angevin distributions and their discrete approximations}.
\newblock {\em Bernoulli}, 2(4):341 -- 363, 1996.

\bibitem{roberts_1996_aperiodic}
G.~O. Roberts and R.~L. Tweedie.
\newblock Geometric convergence and central limit theorems for multidimensional {H}astings and {M}etropolis algorithms.
\newblock {\em Biometrika}, 83(1):95--110, 1996.

\bibitem{Roy2022}
V.~Roy and L.~Zhang.
\newblock Convergence of position-dependent {MALA} with application to conditional simulation in {GLMM}s.
\newblock {\em Journal of Computational and Graphical Statistics}, 32(2):501–512, October 2022.

\bibitem{Rudolph1994}
G.~Rudolph.
\newblock Convergence analysis of canonical genetic algorithms.
\newblock {\em IEEE Transactions on Neural Networks}, 5(1):96–101, 1994.

\bibitem{Sen2025}
N.~Sen, I.~Modak, B.~C. Giri, and S.~Bardhan.
\newblock Vendor managed inventory under price-and-stock-dependent demand, buyer’s space limitations and preservation investment.
\newblock {\em IMA Journal of Management Mathematics}, February 2025.

\bibitem{Sharma2025}
N.~Sharma, M.~Jain, and D.~K. Sharma.
\newblock Metaheuristic optimization for multi-item supply chains with price indices and advance payment.
\newblock {\em IMA Journal of Management Mathematics}, April 2025.

\bibitem{Syswerda1991}
G.~Syswerda.
\newblock {\em A Study of Reproduction in Generational and Steady-State Genetic Algorithms}, page 94–101.
\newblock Elsevier, 1991.

\bibitem{Vehtari2021}
A.~Vehtari, A.~Gelman, D.~Simpson, B.~Carpenter, and P.-C. B\"{u}rkner.
\newblock Rank-normalization, folding, and localization: An improved rˆ for assessing convergence of MCMC (with discussion).
\newblock {\em Bayesian Analysis}, 16(2), June 2021.

\bibitem{Watanabe2011}
S.~Watanabe and N.~Ikeda.
\newblock {\em Stochastic Differential Equations and diffusion processes}.
\newblock Elsevier, Amsterdam, Netherlands, August 2011.

\bibitem{Xifara2014}
T.~Xifara, C.~Sherlock, S.~Livingstone, S.~Byrne, and M.~Girolami.
\newblock {L}angevin diffusions and the {M}etropolis-adjusted {L}angevin algorithm.
\newblock {\em Statistics \& Probability Letters}, 91:14–19, August 2014.

\bibitem{ern1985}
V.~Černý.
\newblock Thermodynamical approach to the traveling salesman problem: {A}n efficient simulation algorithm.
\newblock {\em Journal of Optimization Theory and Applications}, 45(1):41–51, January 1985.

\end{thebibliography}

\appendix
\section{Proof of Theorem~\ref{theorem:invariant_subspace}}\label{sec:proof_theorem_invariant_subspace}
\begin{proof}:
    Since the coefficients are smooth on $U^\circ$, the SDE has continuous solutions up to hitting time $T$. As such, it suffices to prove that no individual slack $s_i(X_t)=b_i-a_i\cdot X_t$ can reach zero in finite time.

    Fix a face $i$, and write $S_t = s_i(X_t)$. Then the SDE for $S_t$ is given by
    \begin{equation}\label{eq:sde_slack}
        dS_t\ =\ \left(a_i^\top C(X_t)\nabla f(X_t)-a_i^\top (\nabla\cdot C)(X_t)\right)dt\ -\ \sqrt{2}a_i^\top C(X_t)^{1/2}dW_t,
    \end{equation}
    for $0\leq t\leq T$, and $S_0=s_i(X_0)>0$.

    \textbf{Part 1}: We first produce bounds for $H(x)$ and $C(x)$. Since $U$ is compact, for $x\in U^\circ$,
    \begin{equation*}
        0\ <\ s_i(x)\ \leq\ \overline{s}_i\ :=\ \max_{y\in U}s_i(y)\ <\ \infty\ \implies\ H(x)\ :=\ \sum_{i=1}^K\frac{a_i a_i^\top}{s_i(x)^2}\ \succcurlyeq\ \sum_{i=1}^K\frac{a_i a_i^\top}{\overline{s}_i^2}\ =:\ H_*.
    \end{equation*}
    In particular, since $U$ is compact with nonempty interior, the normals $a_1,\ldots, a_K$  span $\mathbb{R}^d$, so $H_*$ is positive definite. Thus $C(x) \preccurlyeq H_*^{-1} \preccurlyeq MI$, for some constant $M>0$, and all $x\in U^\circ$. Furthermore, since 
    \begin{equation*}
        H(x)-\frac{a_i a_i^\top}{s_i(x)^2}\ =\ \sum_{j\neq i}\frac{a_j a_j^\top}{s_j(x)^2}\ \succcurlyeq\ 0,\ \implies\ \frac{a_i a_i^\top}{s_i(x)^2}\ \preccurlyeq\ H(x).
    \end{equation*}
    Since $H(x)$ is symmetric and positive definite, we can conjugate by $H(x)^{-1/2}$
    \begin{equation*}
        H(x)^{-1/2}\frac{a_i a_i^\top}{s_i(x)^2}H(x)^{-1/2}\ \preccurlyeq\ I\ \implies\ \frac{a_i^\top C(x)a_i}{s_i(x)^2}\ \leq\ 1.
    \end{equation*}
    Thus, $a_i^\top C(x)a_i\ \leq\ s_i(x)^2$, and by Cauchy-Schwarz in the $C(x)$-inner product, $|a_i^\top C(x)a_j|\leq s_i(x)s_j(x)$.

    \textbf{Part 2}: Using these bounds, we show that the diffusion and drift terms for \eqref{eq:sde_slack} are of order $s_i(x)$. For the diffusion term, since $|a_i^\top C(x)a_i|\leq s_i(x)^2$, it follows that $|a_i^\top C(x)^{1/2}|\leq s_i(x)$. For the drift, since $C(x)\preccurlyeq MI$ and $\nabla f$ is continuous with $U$ compact, $\|\nabla f(x)\|_\infty \leq G$ for all $x\in U$. Thus,
    \begin{equation*}
        |a_i^TC(x)\nabla f(x)|\ \leq \ \sqrt{a_i^TC(x)a_i}\sqrt{\nabla f(x)^TC(x)\nabla f(x)}\ \leq\ s_i(x)\sqrt{M}G.
    \end{equation*}

    Using the formula for $\nabla\cdot C$, we have
    \begin{equation*}
        a_i^\top(\nabla\cdot C)=-2\sum_{j=1}^K\frac{(a_j^TCa_j)(a_j^TCa_i)}{s_j^3}.
    \end{equation*}
    Hence,
    \begin{equation*}
        |a_i^T(\nabla\cdot C)(x)|\ \leq\ 2\sum_{j=1}^K\frac{s_j(x)^2s_j(x)s_i(x)}{s_j(x)^3}\ =\ 2K s_i(x).
    \end{equation*}
    
    \textbf{Part 3}: We now write \eqref{eq:sde_slack} in terms of a geometric Brownian motion 
    \begin{equation*}
        dS_t\ =\ S_t \mu_i(x)\, dt\ +\ S_t\zeta_i(X_t)^\top\, dW_t,
    \end{equation*}
    where 
    \begin{equation*}
        \mu_i(x)\ =\ \frac{a_i^\top C(x)\nabla f(x) - a_i^\top (\nabla\cdot C)(x)}{s_i(x)},\quad\text{and}\quad \zeta_i(x)\ =\ -\sqrt{2}\frac{C(x)^{1/2}a_i}{s_i(x)}\in\mathbb{R}^d.
    \end{equation*}
    Since the numerator of $\mu_i(x)$ and $\zeta_i(x)$ are of order $s_i(x)$, $|\mu_i(x)|\leq L_i$ for some constant $L_i$ and $|\zeta_i(x)|^2\leq 2$, for $x\in U^\circ$. Applying It\^o's formula to $\log (S_t)$ on $[0,T)$ yields
    \begin{equation*}
        d\log S_t\ =\ \left(\mu_i(X_t) - \frac{1}{2}|\zeta_i(X_t)|^2\right)\, dt\ +\ \zeta_i(X_t)^\top dW_t.
    \end{equation*}
    Thus, for every $t\geq 0$,
    \begin{equation*}
        S_{t\wedge T}\ =\ S_0\exp\left(\int^{t\wedge T}_0 \left(\mu_i(X_r) - \frac{1}{2}|\zeta_i(X_r)|^2\right)\, dr\ +\ \int^{t\wedge T}_0\zeta_i(X_r)^\top dW_r\right).
    \end{equation*}
    Since the coefficients are bounded, both integrals are finite for every finite $t$. Hence, $S_{t\wedge T}>0$ a.s. for every finite $t$.
\end{proof}

\end{document}